\documentclass[aps, pra, onecolumn,notitlepage, amsmath, amssymb, superscriptaddress, nofootinbib]{revtex4-1}

\makeatletter
\def\p@subsection{}
\def\p@subsubsection{}
\makeatother

\usepackage[utf8]{inputenc}
\usepackage{graphicx}
\usepackage{units}
\usepackage{color}
\usepackage[pdftex, colorlinks=true, linkcolor=myblue, citecolor=myblue,
urlcolor=myblue]{hyperref}
\usepackage{times}
\usepackage{comment}
\usepackage{graphicx}
\usepackage{amsmath, calc}
\usepackage{mathrsfs}
\usepackage{MnSymbol}
\usepackage{amsfonts}
\usepackage{amssymb}
\usepackage{bm}
\usepackage{bbm}
\usepackage{color}
\usepackage{array}
\usepackage{units}
\usepackage{mathrsfs,amsmath}

\definecolor{myblue}{rgb}{0,0,1}
\definecolor{myred}{rgb}{1,0,0}

\newcommand{\ket}[1]{|#1\rangle}

\newcommand{\kettwo}[1]{||#1\rangle \rangle}

\begin{document}
\title{Quantum topology in the ultrastrong coupling regime}

\author{C.~A.~Downing}
\email{c.a.downing@exeter.ac.uk} 
\affiliation{Department of Physics and Astronomy, University of Exeter, Exeter EX4 4QL, United Kingdom}

\author{A.~J.~Toghill}
\affiliation{Department of Physics and Astronomy, University of Exeter, Exeter EX4 4QL, United Kingdom}

\date{\today}

\begin{abstract}
\noindent \textbf{Abstract}\\
The coupling between two or more objects can generally be categorized as strong or weak. In cavity quantum electrodynamics for example, when the coupling strength is larger than the loss rate the coupling is termed strong, and otherwise it is dubbed weak. Ultrastrong coupling, where the interaction energy is of the same order of magnitude as the bare energies of the uncoupled objects, presents a new paradigm for quantum physics and beyond. As a consequence profound changes to well established phenomena occur, for instance the ground state in an ultrastrongly coupled system is not empty but hosts virtual excitations due to the existence of processes which do not conserve the total number of excitations. The implications of ultrastrong coupling for quantum topological systems, where the number of excitations are typically conserved, remain largely unknown despite the great utility of topological matter. Here we reveal how the delicate interplay between ultrastrong coupling and topological states manifests in a one-dimensional array. We study theoretically a dimerized chain of two-level systems within the ultrastrong coupling regime, where the combined saturation and counter-rotating terms in the Hamiltonian are shown to play pivotal roles in the rich, multi-excitation effective bandstructure. In particular, we uncover unusual topological edge states, we introduce a flavour of topological state which we call an anti-edge state, and we reveal the remarkable geometric-dependent renormalizations of the quantum vaccum. Taken together, our results provide a route map for experimentalists to characterize and explore a prototypical system in the emerging field of ultrastrong quantum topology.
\end{abstract}


\maketitle


\noindent \textbf{Introduction}\\
The beauty of topology continues to fascinate scientists from an increasing diversity of fields, including more recently the quantum light and matter community~\cite{Chang2018}. The versatility of modern systems from the nanophotonic~\cite{Lu2014, Ozawa2019, Smirnova2020} to the magnonic~\cite{McClarty2021} to the ultracold atomic~\cite{Cooper2019} has significantly enrichened contemporary topological physics. Indeed, quantum topology is advancing at an impressive rate both from a fundamental point of view, including the creation of topological sources of quantum light~\cite{Mittal2018} and biphoton states~\cite{Blanco2018}, and from the perspective of applications, for example the recent development of topological lasers~\cite{Bandres2018} and chiral quantum optical devices~\cite{Barik2018}. 

\begin{figure}[tb]
 \includegraphics[width=1.0\linewidth]{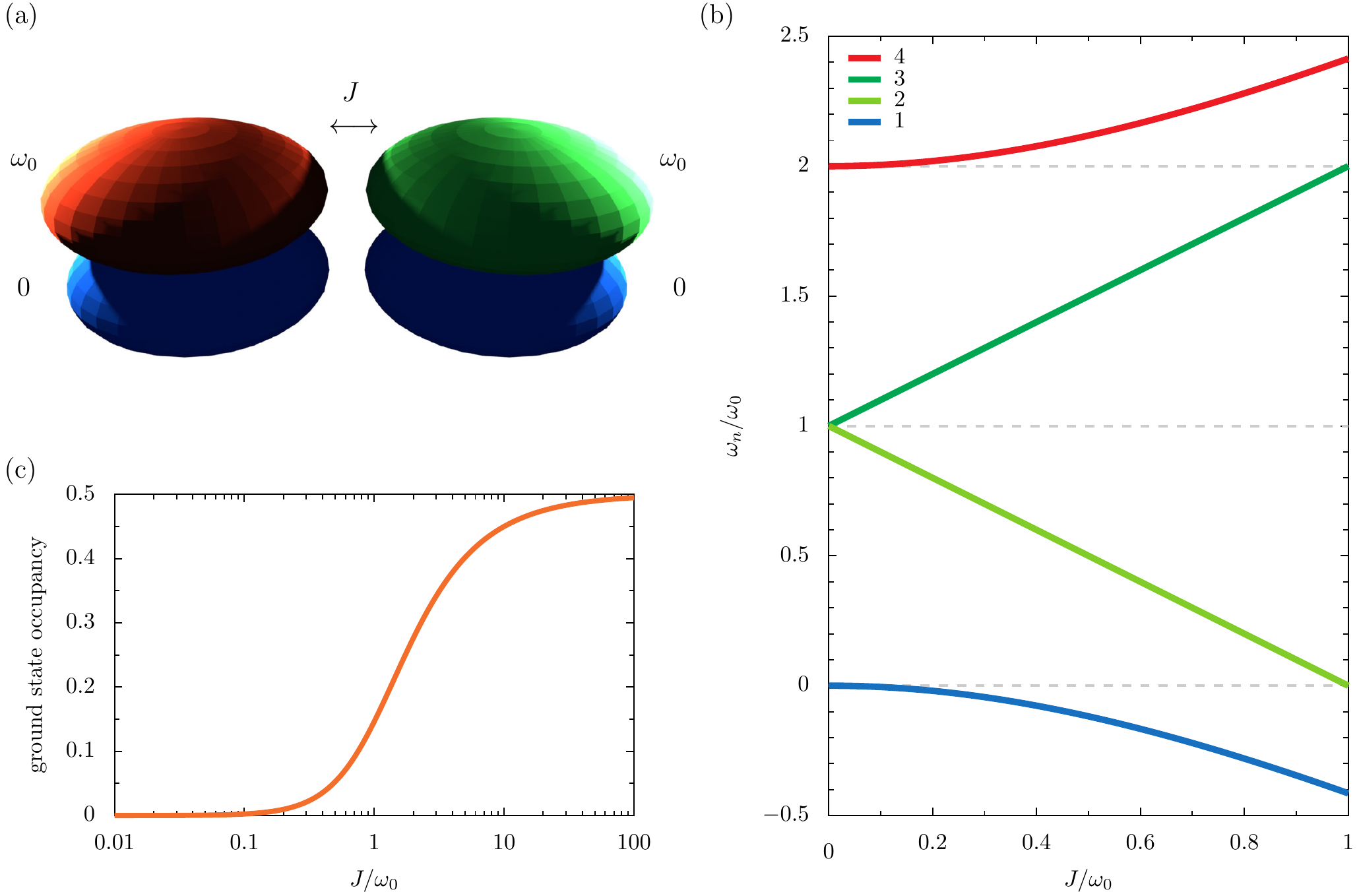}
 \caption{ \textbf{Ultrastrong coupling in the dimer.} Panel (a): a sketch of a pair of two-level systems, each of bare transition frequency $\omega_0$, with the qubit-qubit coupling strength $J$ [cf. Eq.~\eqref{eq:dimerrrr}]. Panel (b): the eigenfrequencies $\omega_n$, as a function of $J$ [cf. Eq.~\eqref{eq:eigenfrequencies}]. Dashed grey lines: the transition frequencies in the uncoupled limit. Panel (c): the occupancy of the ground state $\kettwo{\psi_1}$ as a function of $J$ [cf. Eq.~\eqref{eq:sda444}].}
 \label{dimerr}
\end{figure}

A standard approximation employed within quantum physics is the so-called rotating wave approximation (RWA), whereby all non-resonant terms in the Hamiltonian are discarded as negligible~\cite{Allen1975}. This celebrated approximation breaks down in the ultrastrong coupling regime~\cite{Kockum2019, Forn2019, Boite2020}, where the coupling strength becomes comparable to the bare transition frequencies of the uncoupled subsystems. The coupling is no longer perturbative, such that the non-resonant counter-rotating (C-R) terms must be included in any proper treatment~\cite{Peropadre2013, Naether2014, Sanchez2014, Calvo2020, Roman2020,Gonzalez2021}. Importantly, these C-R terms permit processes which violate the conservation of the number of excitations, so that the intuitively empty ground state instead bubbles with virtual excitations. The impact of ultrastrong coupling on quantum topological matter~\cite{Asboth2016, Haldane2017}, which usually draws upon a single particle picture, presents an interesting puzzle.

Ultrastrong coupling has been observed in a stream of pioneering experiments across several platforms, from $LC$ resonators magnetically coupled to superconducting qubits~\cite{Forn2010}, to superconducting artificial atoms coupled to on-chip cavities~\cite{Niemczyk2010}, to superconducting artificial atoms coupled to the electromagnetic continuum of a one-dimensional waveguide~\cite{Forn2017}, to arrays of plasmonic nanoparticles~\cite{Mueller2020}. Therefore, an expansion of the field to include topological considerations - so-called ultrastrong quantum topology - is a natural progression which promises a myriad of theoretical and experimental curiosities. Early theoretical efforts have thus far focussed on the influence of the geometric phase and other topological features of the quantum Rabi model~\cite{Liu2021, Ying2022}.

A prototypical model showcasing ultrastrong coupling is a pair of coupled two level systems (2LSs). As sketched in Fig.~\ref{dimerr}~(a), we consider the transition frequency of each 2LS to be $\omega_0$, and the coupling strength to be $J > 0$. The Hamiltonian $\hat{H}$, including the C-R terms, thus reads (after setting $\hbar = 1$)~\cite{Allen1975, Ficek2002}
\begin{equation}
\label{eq:dimerrrr}
 \hat{H} = \omega_0  \left( \sigma_1^{\dagger} \sigma_1 + \sigma_2^{\dagger} \sigma_2 \right) + J \left( \sigma_1 + \sigma_1^{\dagger} \right) \left( \sigma_{2} + \sigma_{2}^{\dagger} \right),
\end{equation}
with the raising (lowering) operator $\sigma_n^\dagger$ ($\sigma_n$) describing the $n$-th 2LS. Within the RWA, the terms $\propto \sigma_1 \sigma_2$ and $\propto \sigma_1^{\dagger} \sigma_2^{\dagger}$ in Eq.~\eqref{eq:dimerrrr} are discarded and the four resulting eigenfrequencies $\omega_n'$ are simply: $\omega_4' = 2\omega_0$; the hybridized levels $\omega_3' = \omega_0+J$ and $\omega_2' = \omega_0-J$; and the ground state $\omega_1' = 0$. In terms of the bare states in the occupation number representation, that is $\ket{0, 0}$, $\ket{1, 0}$, $\ket{0, 1}$, and $\ket{1, 1}$, one may find the corresponding eigenstates $\ket{\psi_n}$. The extremities of the energy ladder are associated with the doubly occupied eigenstate, which we denote $\ket{\psi_4} = \ket{1, 1}$, and the wholly unoccupied ground state, which we label $\ket{\psi_1} = \ket{0, 0}$. The intermediate, singly-occupied eigenstates $\ket{\psi_3} = \left( \ket{1, 0} + \ket{0, 1} \right)/\sqrt{2}$ and $\ket{\psi_2} = \left( \ket{1, 0} - \ket{0, 1} \right)/\sqrt{2}$ are superpositions due to the nonzero coupling, and are separated in energy by the splitting $2J$~\cite{Downing2019, Downing2020}.

When considering the ultrastrong coupling regime, diagonalizing the full Hamiltonian of Eq.~\eqref{eq:dimerrrr} leads to the exact eigenfrequencies $\omega_n$, which are given by~\cite{Decordi2017}
\begin{subequations}
\label{eq:eigenfrequencies}
\begin{align}
 \omega_{4} &=  \omega_0 + \sqrt{ \omega_0^2 + J^2}. \label{eq:eigenfsdrequenciesC}\\
 \omega_{3} &= \omega_0 + J, \label{eq:eigenfrequsdenciesB} \\
  \omega_{2} &= \omega_0 - J, \label{eq:eigsdenfresdquenciesBB} \\
 \omega_{1} &= \omega_0 - \sqrt{ \omega_0^2 + J^2}. \label{eq:eigensdfrequenciesA} 
  \end{align}
\end{subequations}
Notably, the C-R terms $\propto \sigma_1 \sigma_2$ and $\propto \sigma_1^{\dagger} \sigma_2^{\dagger}$ in Eq.~\eqref{eq:dimerrrr} link the zero-excitation and two-excitation sectors, breaking particle number conservation. Consequently, the highest and lowest rungs of the energy ladder are renormalized from $\omega_4' = 2\omega_0$ and $\omega_1' = 0$ to the $J$-dependent eigenfrequencies $\omega_4$ and $\omega_1$ respectively [cf. Eq.~\eqref{eq:eigenfsdrequenciesC} and Eq.~\eqref{eq:eigensdfrequenciesA}]. In Fig.~\ref{dimerr}~(b), we plot the four eigenfrequencies $\omega_n$ of Eq.~\eqref{eq:eigenfrequencies} as a function of the coupling strength $J$, showcasing the drastic reconstruction of half of the energy ladder in the ultrastrong coupling regime. In particular, there are divergences from the bare transition frequencies $0$ and $2\omega_0$, as marked by the dashed grey lines. The exact eigenstates associated with Eq.~\eqref{eq:eigenfrequencies}, where we use the notation $\kettwo{\psi_n}$ for when the C-R terms are considered, read
\begin{subequations}
\label{eq:eigenstates}
\begin{align}
 \kettwo{\psi_4} &= \frac{1}{\sqrt{ \omega_{1}^2 + J^2 }} \left( J \ket{1, 1} - \omega_{1} \ket{0, 0} \right), \label{eq:sda111} 
\\
  \kettwo{\psi_3} &= \frac{1}{\sqrt{2}} \left( \ket{1, 0} + \ket{0, 1} \right), \label{eq:sda222}  \\
    \kettwo{\psi_2} &= \frac{1}{\sqrt{2}} \left( \ket{1, 0} - \ket{0, 1} \right), \label{eq:sda333}  \\
  \kettwo{\psi_1} &= \frac{1}{\sqrt{ \omega_{4}^2 + J^2 }} \left( \omega_{4} \ket{0, 0} -  J \ket{1, 1} \right), \label{eq:sda444}  
 \end{align}
\end{subequations}
where the frequencies $\omega_{1}$ and $\omega_{4}$ are defined in Eq.~\eqref{eq:eigenfrequencies}. Clearly, the mixed particle number eigenstates of Eq.~\eqref{eq:sda111} and Eq.~\eqref{eq:sda444} present a new paradigm for coupled systems. One immediate implication is that the ground state $\kettwo{\psi_1}$ is no longer trivially empty, $\kettwo{\psi_1} \ne \ket{0, 0}$. As shown in Fig.~\ref{dimerr}~(c), the occupancy of the proper ground state $\kettwo{\psi_1}$ increases with coupling strength $J$. When $J \ll \omega_0$, the pure state $\kettwo{\psi_1} \simeq \ket{0, 0}$ means that the ground state is wholly unoccupied, while in the extreme limiting case of $J \gg \omega_0$ the maximally mixed state of $\kettwo{\psi_1} \simeq \left( \ket{0, 0} - \ket{1, 1} \right)/\sqrt{2}$ is essentially reached, such that the ground state $\kettwo{\psi_1}$ is half-occupied and half-unoccupied.

\begin{figure}[tb]
 \includegraphics[width=0.7\linewidth]{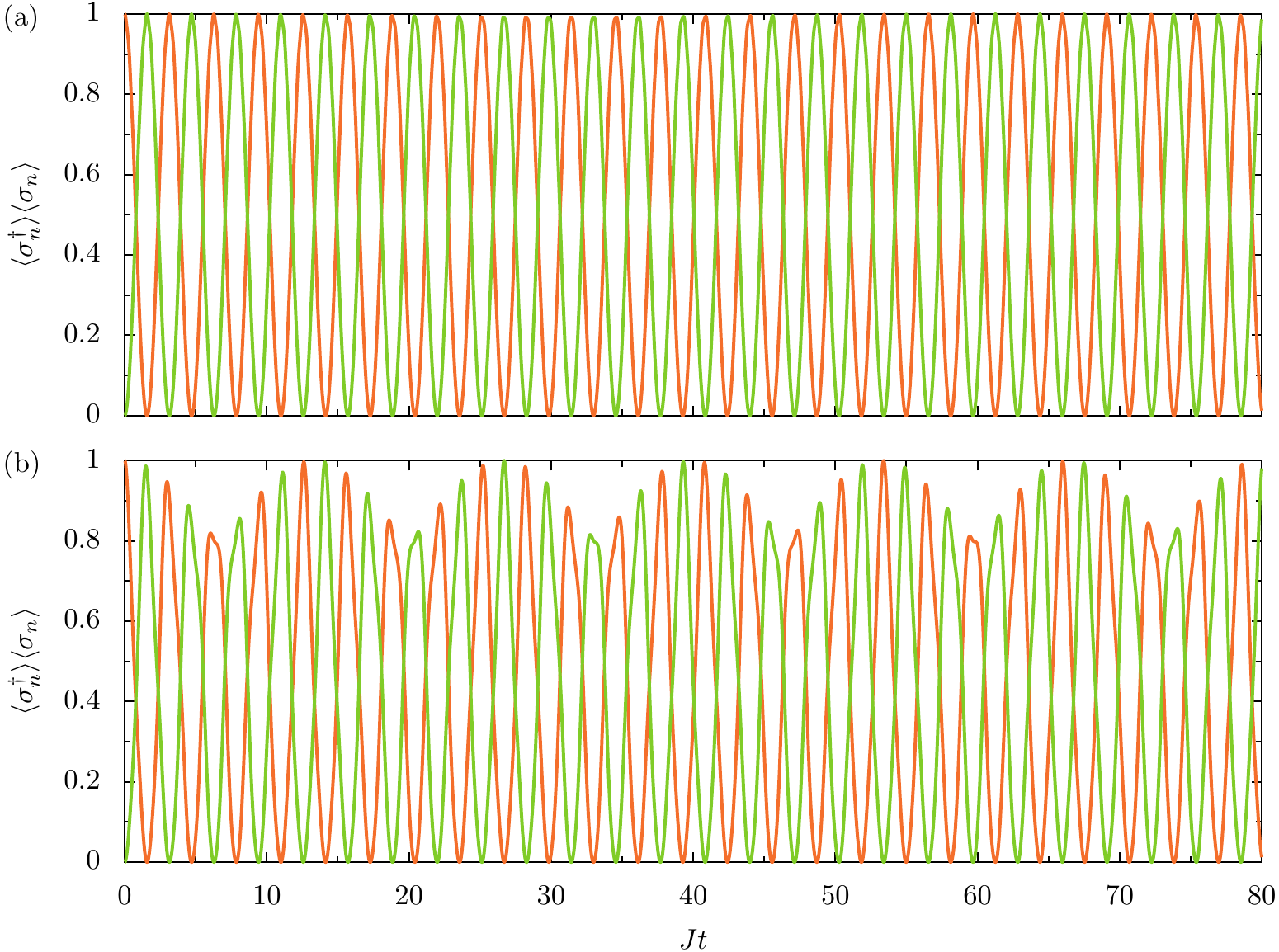}
 \caption{ \textbf{Mean correlations in the dimer.}  The evolution of $\langle  \sigma_n^\dagger \rangle \langle  \sigma_n \rangle$ as a function of time $t$ for the $n$-th 2LS, in units of the inverse coupling strength $J^{-1}$ [cf. Eq.~\eqref{eq:ppiio} with Eq.~\eqref{eq:ghjghjgj}]. Orange lines: $n = 1$. Green lines: $n = 2$. Panel (a): the strong coupling regime, where the bare transition frequency $\omega_0 = 10 J$. Panel (b): the ultrastrong coupling regime, where $\omega_0 = 2 J$.}
 \label{mea1}
\end{figure}

The impact of the ultrastrong coupling can also be seen in the mean correlations of the dimer. For example, the population-like quantity $\langle  \sigma_n^\dagger \rangle \langle  \sigma_n \rangle$ reads (see the Supplementary Information)
 \begin{subequations}
\label{eq:ppiio}
\begin{alignat}{5}
 \langle  \sigma_1^\dagger \rangle \langle  \sigma_1 \rangle &=  f\left( \tilde{\omega}_0, t \right) \: \cos^2 \left( J t \right), \\
  \langle  \sigma_2^\dagger \rangle \langle  \sigma_2 \rangle &=  f\left( \tilde{\omega}_0, t \right) \: \sin^2 \left( J t \right),
\end{alignat}
\end{subequations}
where the auxiliary function $f\left( \tilde{\omega}_0, t \right)$, which accounts for the ultrastrong coupling, is given by
\begin{equation}
\label{eq:ghjghjgj}
f\left( \tilde{\omega}_0, t \right) = \cos^2 \left(  \tilde{\omega}_0 t \right) + \left( \frac{\omega_0}{\tilde{\omega}_0} \right)^2 \sin^2 \left(  \tilde{\omega}_0 t \right),
\end{equation}
with the renormalized transistion frequency $\tilde{\omega}_0 = \sqrt{ \omega_0^2 + J^2}$. In the strong coupling limit of $J \ll \omega_0$, the function $f\left( \tilde{\omega}_0, t \right) \simeq 1$ and Eq.~\eqref{eq:ppiio} approaches the simple trigonometric results $\langle  \sigma_1^\dagger \rangle \langle  \sigma_1 \rangle \simeq  \cos^2 \left( J t \right)$ and $\langle  \sigma_2^\dagger \rangle \langle  \sigma_2 \rangle \simeq  \sin^2 \left( J t \right)$, which recover the expressions found after neglecting the C-R terms in the Hamiltonian of Eq.~\eqref{eq:dimerrrr}. Otherwise, Eq.~\eqref{eq:ppiio} presents nontrivalities. In Fig.~\ref{mea1} the evolution of the population-like cycles of $\langle  \sigma_n^\dagger \rangle \langle  \sigma_n \rangle$ are displayed as a function of time $t$ using Eq.~\eqref{eq:ppiio}, where the orange and green lines represent the first and second 2LS respectively [cf. Fig.~\ref{dimerr}~(a)]. Panel (a), where $J = \omega_0/10$, shows the typical strong coupling regime result: periodic cycles of essentially unity amplitude, which are well described by trigonometric functions in the dimensionless quantity $J t$. The ultrastrong coupling regime $\omega_0 \sim J$ sees the renormalized transition frequency $\tilde{\omega}_0$ influence the amplitude of the cycles following Eq.~\eqref{eq:ghjghjgj}. This is typified by panel (b), where $J = \omega_0/2$ and a characteristic wave packet is observed oscillating in time. The C-R terms also also impact the entanglement properties of the dimer, as discussed in the Supplementary Information.

The consequences of ultrastrong coupling for quantum topology is mostly unchartered terrain. We will start to explore it via the celebrated Su-Schrieffer-Heeger (SSH) topological array model~\cite{Cooper2019, Su1979, Asboth2016}, which is formed by a chain of dimers like the one described in Fig.~\ref{dimerr}.\\

\begin{figure*}[tb]
 \includegraphics[width=1.0\linewidth]{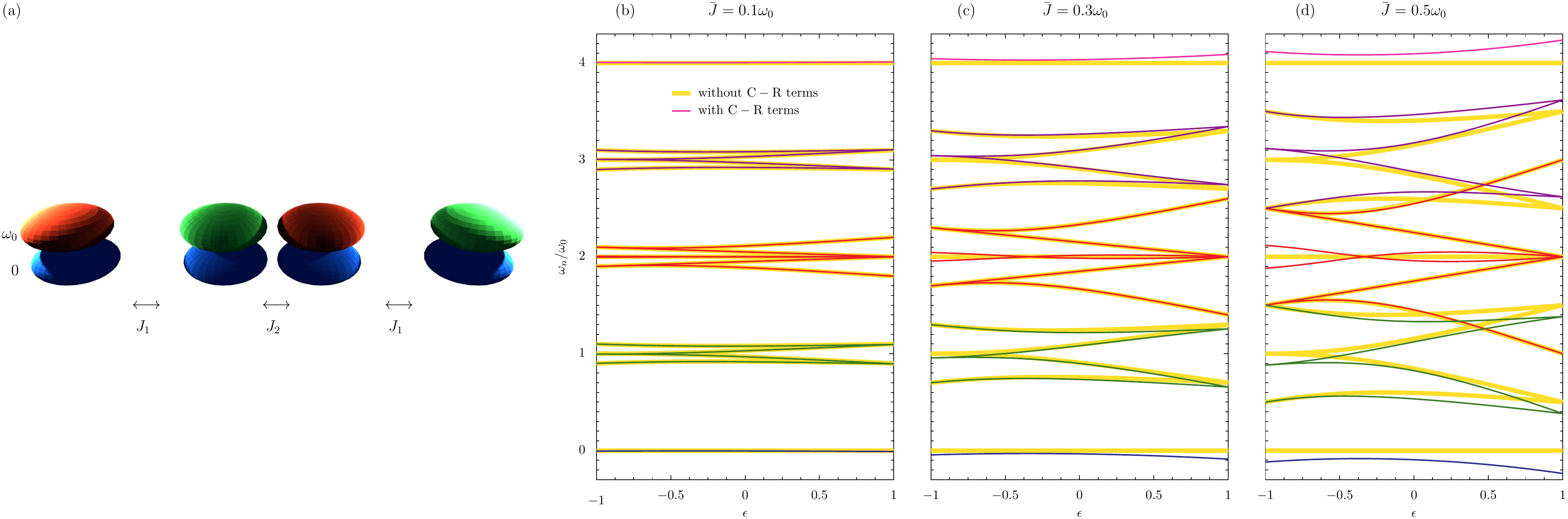}
 \caption{ \textbf{Ultrastrong coupling in the dimerized chain.} Panel (a): a sketch of a bipartite array pair of two-level systems, each of bare transition frequency $\omega_0$, with alternating coupling strengths $J_1$ and $J_2$ along the chain of size $N = 4$ [cf. Eq.~\eqref{eq:hammyy}]. Panels (b, c, d): the eigenfrequencies as a function of the dimerization parameter $\epsilon$, for increasing coupling strength $\bar{J}$ across the panels [cf. Eq.~\eqref{eq:xzzcxzczc}]. Thin colored lines: the eigenfrequencies $\omega_n$, found by diagonalizing the full Hamiltonian of Eq.~\eqref{eq:hammyy}. Thick yellow lines: $\omega_n^\prime$, found upon neglecting the C-R terms. }
 \label{combo}
\end{figure*}


\noindent \textbf{Results}\\
Let us consider a dimerized chain of 2LSs, each of bare transition frequency $\omega_0$, with alternating coupling strengths $J_1 > 0$ and $J_2 > 0$ along the formed one-dimensional lattice~\cite{ Asboth2016}. The arrangement is sketched in Fig.~\ref{combo}~(a) for a short example chain of four 2LSs. For a general chain of size $N$, the Hamiltonian $\hat{H}$ reads [cf. Eq.~\eqref{eq:dimerrrr}]
\begin{equation}
\label{eq:hammyy}
 \hat{H} = \omega_0  \sum_{n=1}^{N} \sigma_n^{\dagger} \sigma_n + J_1 \sum_{n=1}^{\lfloor{\frac{N}{2}}\rfloor } \left( \sigma_{2n} + \sigma_{2n}^{\dagger} \right) \left( \sigma_{2n-1} + \sigma_{2n-1}^{\dagger} \right)
+ J_2 \sum_{n=1}^{\lfloor{\frac{N-1}{2}}\rfloor } \left( \sigma_{2n} + \sigma_{2n}^{\dagger} \right) \left( \sigma_{2n+1} + \sigma_{2n+1}^{\dagger} \right),
\end{equation}
where the floor function $\lfloor{...}\rfloor$ ensures Eq.~\eqref{eq:hammyy} holds for both even and odd values of the integer $N$. The model encapsulated by Eq.~\eqref{eq:hammyy} has an equivalent $2^N \times 2^N$ matrix representation, and in the RWA there are well-defined sectors corresponding to the conserved number of excitations $\mathcal{N}$. The sectors follow Pascal’s triangle, where the binomial coefficient $N!/ \mathcal{N}!/(N-\mathcal{N})!$ counts the number of eigenstates in each excitation sector $\mathcal{N}$ (see the Supplementary Information). For a chain of $N = 4$ 2LSs, the $2^4 = 16$-dimensional Hilbert space is distributed with $\{1, 4, 6, 4, 1\}$ eigenstates in the $\mathcal{N} = \{ 0, 1, 2, 3, 4\}$ excitation sectors respectively (when working in the RWA). That is, there is a single excitation-less ($\mathcal{N} = 0$) ground state, four single-excitation ($\mathcal{N} = 1$) states, six two-excitation ($\mathcal{N} = 2$) states and so on. Two principal parameters govern the intrinsic physics of the model defined by Eq.~\eqref{eq:hammyy},
\begin{equation}
\label{eq:xzzcxzczc}
  \epsilon = \frac{J_1-J_2}{\bar{J}},
  \quad\quad\quad\quad\quad\quad\quad\quad\quad\quad\quad
  \bar{J} =  J_1 + J_2,
\end{equation}
namely $\epsilon$, the dimerization parameter which records the inherent geometry of the bipartite chain; and $\bar{J}$, the chain coupling strength. The latter quantity tracks the importance of the C-R terms like $\propto \sigma_l \sigma_m$ and $\propto \sigma_l^{\dagger} \sigma_m^{\dagger}$ in Eq.~\eqref{eq:hammyy} via the dimensionless ratio $\bar{J}/\omega_0$. The inverse relations for Eq.~\eqref{eq:xzzcxzczc} provide forms of the alternating coupling strengths $J_1 = \left( 1 + \epsilon \right)  \bar{J} / 2$ and $J_2 = \left( 1 - \epsilon \right)  \bar{J} / 2$.

\begin{figure*}[tb]
 \includegraphics[width=0.85\linewidth]{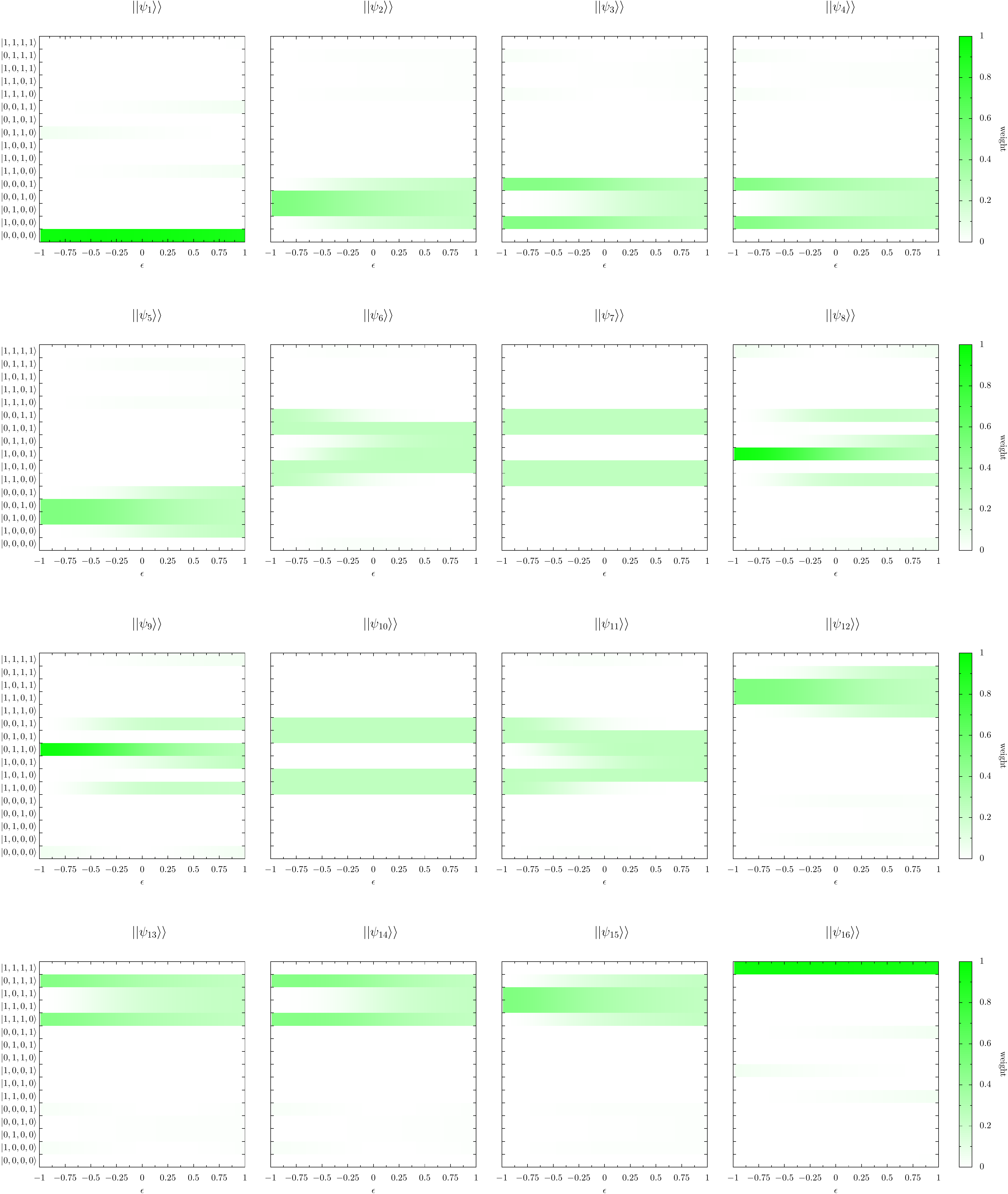}
 \caption{ \textbf{Eigenstates in the dimerized chain.} The probability densities of the eigenstates $\kettwo{\psi_n}$ in the ultrastrong coupling regime, in terms of the bare states $\ket{i, j, k, l}$, as a function of the dimerization $\epsilon$ [cf. Eq.~\eqref{eq:xzzcxzczc}]. In the figure, the coupling strength $\bar{J} = 0.5 \omega_0$, and the chain is of size $N = 4$ (corresponding to a $2^4=16$ dimensional Hilbert space). The presented eigenstates correspond to the eigenfrequencies of Fig.~\ref{combo}~(d). }
 \label{ssheig5}
\end{figure*}

\begin{figure*}[tb]
 \includegraphics[width=1.0\linewidth]{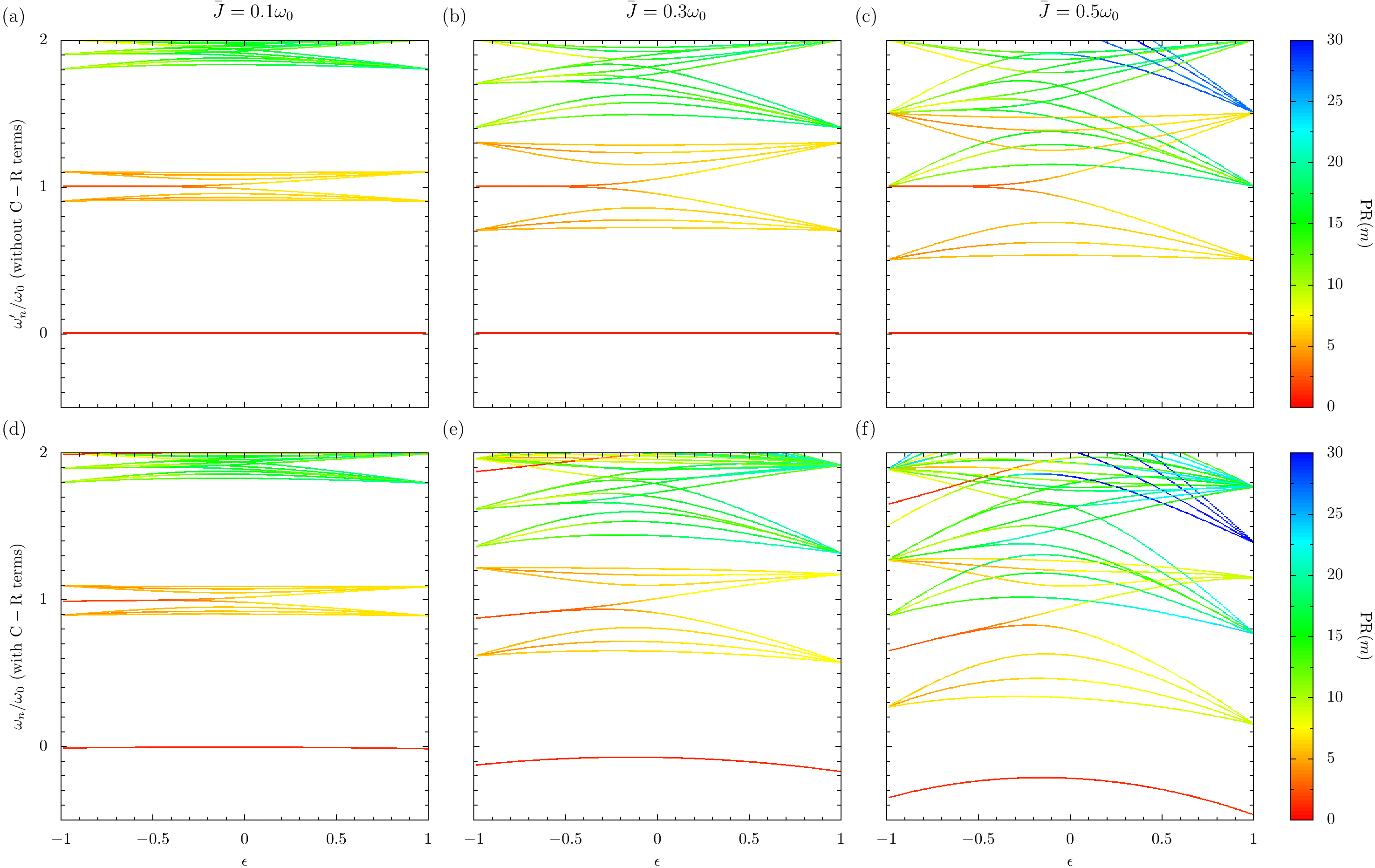}
 \caption{ \textbf{Topology in the ultrastrong coupling regime.} Eigenfrequencies as a function of the dimerization parameter $\epsilon$, with increasing coupling strength $\bar{J}$ in each column [cf. Eq.~\eqref{eq:xzzcxzczc}]. Colour bars: the participation ratio $\mathrm{PR}(n)$ of each eigenstate. Panels (a, b, c): the results without taking counter-rotating terms in the Hamiltonian of Eq.~\eqref{eq:hammyy} into account, namely the RWA eigenfrequencies $\omega_n^\prime$ and RWA eigenstates $\ket{\psi_n}$. Panels (d, e, f): the full eigenfrequencies $\omega_n$ and eigenstates $\kettwo{\psi_n}$ are used. In the figure, the chain is of size $N = 8$ (corresponding to a $2^8=256$ dimensional Hilbert space) and all data above $2\omega_0$ is cut.   }
 \label{pr33}
\end{figure*}

The results of diagonalizing Eq.~\eqref{eq:hammyy} are shown in Fig.~\ref{combo}~(b, c, d) for a chain of $N = 4$ 2LSs, which is represented in Fig.~\ref{combo}~(a). The panels (b, c, d) display the eigenfrequencies as a function of the dimerization $\epsilon$, with increasing coupling strength $\bar{J}$ across the panels. In panel (b) the chain coupling strength $\bar{J} = 0.1 \omega_0$, such that the $16$ exact eigenfrequencies $\omega_n$ (thin colored lines) closely resemble the results in the RWA (thick yellow lines). Notably, the effective bandstructure with dimerization $\epsilon < 0$ is markedly different from that with $\epsilon > 0$, especially around $\omega_0$ and $3\omega_0$ where states are either present or absent depending upon the dimerization (hinting at the topology of the model). Upon increasing the coupling strength to $\bar{J} = 0.3 \omega_0$ in panel (c), the influence of the C-R terms becomes visible, particularly around $0$ and $4\omega_0$, where there are significant deviations from the RWA results [as may have been anticipated from the dimer results of Fig.~\ref{dimerr}~(b), where the $0$ and $2$ excitation sectors became linked]. Finally, in panel (d) $\bar{J} = 0.5 \omega_0$, such that the impact of entering the ultrastrong coupling regime is highly apparent. It has lead even to crossovers between energy levels supposedly (in the RWA) associated with different numbers of excitations. For example, green (nominally single excitation sector) and red (supposedly two excitation sector) eigenfrequencies overlap when $\epsilon > 0$, as do the red and purple (purportedly three excitation sector) results. However, while there is a significant reconstruction of the energy ladder, the asymmetry about $\epsilon = 0$ in panel (d) remains in a similar fashion to panel (a), hinting that some topological properties may well endure.

The underlying behavior of the eigenstates associated with Fig.~\ref{combo}~(d) is investigated in Fig.~\ref{ssheig5}, where we consider the fidelity of the $16$ quantum states from $\kettwo{\psi_1}$ to $\kettwo{\psi_{16}}$, ordered from lowest to highest in energy (from $\omega_1$ to $\omega_{16}$). That is, we show how the weighting of the underlying bare states $\ket{i, j, k, l}$ changes as a function of the dimerization $\epsilon$, with white signifying zero contribution from the bare state and brighter shades of green denoting an increasingly large overlap (the bare states $\ket{i, j, k, l}$ are marked at the left edge of the first column of panels). The eigenstates labelled $\kettwo{\psi_3}$ and $\kettwo{\psi_4}$ correspond to the states residing near to $\omega_0$ (when $\epsilon < 0$) in Fig.~\ref{combo}~(d). The overlaps of $\kettwo{\psi_3}$ and $\kettwo{\psi_4}$ (found at the right-hand side of the upper row of Fig.~\ref{ssheig5}) demonstrates their transition from being primarily edge states when $\epsilon < 0$, composed of the bare states $\ket{1, 0, 0, 0}$ and $\ket{0, 0, 0, 1}$, to being bulk states extended throughout the four-dimensional single excitation sector when $\epsilon > 0$. An analogous effect may be observed for the eigenstates $\kettwo{\psi_{13}}$ and $\kettwo{\psi_{14}}$, which reside around $3\omega_0$ (when $\epsilon < 0$) in Fig.~\ref{combo}~(d). Now it is a `hole', or absence of excitation, which appears as a novel type of topological anti-edge state, based upon the bare states $\ket{1, 1, 1, 0}$ and $\ket{0, 1, 1, 1}$ (see the two panels on the left-hand side of the bottom row of Fig.~\ref{ssheig5}). These edge state and anti-edge state features are common across chains of an arbitrary size $N$, even in the ultrastrong coupling regime (see the Supplementary Information). Indeed, despite the significant coupling strength of $\bar{J} = 0.5 \omega_0$ the packaging of states into excitation-number-conserving bundles is still has some utility. The influence of the C-R terms on the eigenstates $\kettwo{\psi_n}$ can be seen for example in the bottom right panel of Fig.~\ref{ssheig5}, describing the eigenstate $\kettwo{\psi_{16}}$, where the deviation from the RWA result of $\ket{1, 1, 1, 1}$ arises in the noticeable contributions of the two-excitation sector bare states.

The topology of a dimerized chain of 2LSs in the ultrastrong coupling regime is charted in Fig.~\ref{pr33} for a typical non-short chain. There we plot the eigenfrequencies as a function of the dimerization parameter $\epsilon$, with increasing coupling strength $\bar{J}$ for each column. The top row employs the RWA and as such is applicable within strong coupling, while the bottom row takes the C-R terms into account so that ultrastrong coupling may also be properly described. In the figure, the chain is of size $N = 8$ (corresponding to a $2^8=256$-dimensional Hilbert space) and all data above $2\omega_0$ is cut. The color bar measures the localization of each state via $\mathrm{PR}(n)$, the participation ratio~\cite{Bell1970, Thouless1974, DowningMartin,Pablo} of each eigenstate. In the participation ratio calculation, $\ket{\psi_{n}}$ used in the upper row and $\kettwo{\psi_{n}}$ in the lower row. This localization measure counts how many bare states contribute to the eigenstate, with red marking a low number of states (a signature of potential edge states), and blue a high number of states (a signifier of highly extended states spread out over the entire chain).

In first column of Fig.~\ref{pr33}, where the coupling strength $\bar{J} = 0.1 \omega_0$ effectively ensures strong coupling behavior, the results of panels (a) and (d) are in essence the same. One observes in red the topological edge state at $\omega_0$ for $\epsilon < 0$, which lies in the effective band gap between two collections of extended states in orange-yellow. This midgap state disappears when $\epsilon > 0$, a result resembling standard one-excitation SSH-like models transitioning from the topologically nontrivial to the topologically trivial regime~\cite{Asboth2016}. The lowest energy state $\omega_1^\prime = 0$ in panel (a) [or $\omega_1 \simeq 0$ in panel (d)] is not topological, and it is red simply because it is only composed of the empty state $\ket{0, 0, 0, 0}$  [in panel (d), $\kettwo{\psi_1} \simeq \ket{0, 0, 0, 0}$ holds].

The second column of Fig.~\ref{pr33} sees the coupling strength increased to $\bar{J} = 0.3 \omega_0$. Similar topological features can be found around $\omega_0$, with the red localized state present when $\epsilon < 0$ merging into the effective bands for $\epsilon > 0$. However, now the two effective bands surrounding the edge state are of substantially larger effective bandwidths due to the increased coupling strength $\bar{J}$, such that the nominally two-excitation-sector states in the vicinity of $2\omega_0$ (primarily colored in green) are almost reaching the ostensible one-excitation-sector states around $\omega_0$ (mostly orange-yellow). While the orange-yellow extended states in panel (b) maintain symmetric behavior about $\omega_0$ due to the RWA, the analogous collections of states in panel (d) are highly asymmetric, since the C-R terms guarantee significant couplings to higher rungs of the energy ladder. Most notably, these hybridizations lead to a ground state no longer pinned at $0$ in panel (d).

When $\bar{J} = 0.5 \omega_0$ in the third column of Fig.~\ref{pr33}, there is a drastic reconstruction of the effective bandstructure in panel (f) due to the ultrastrong coupling causing a complete breakdown of particle number conservation. The ground state shows a remarkable warping away from $0$, and there is pronounced effective band crossing around $\omega_0$ for all values of $\epsilon$. However, quite remarkably, the edge state (now located near to $0.7 \omega_0$ when $\epsilon \simeq -1$) persists, as does its highly localized nature (it is red for $\epsilon < 0$ and indeed transforms into an extended state in yellow-green for $\epsilon > 0$) suggesting the translation of topological features deep into the ultrastrong coupling regime. This relative robustness to the effects of the C-R terms in the Hamiltonian presents intriguing perspectives for the existence of localized edge states specifically, and for ultrastrong quantum topology in general.

\begin{figure}[tb]
 \includegraphics[width=1.0\linewidth]{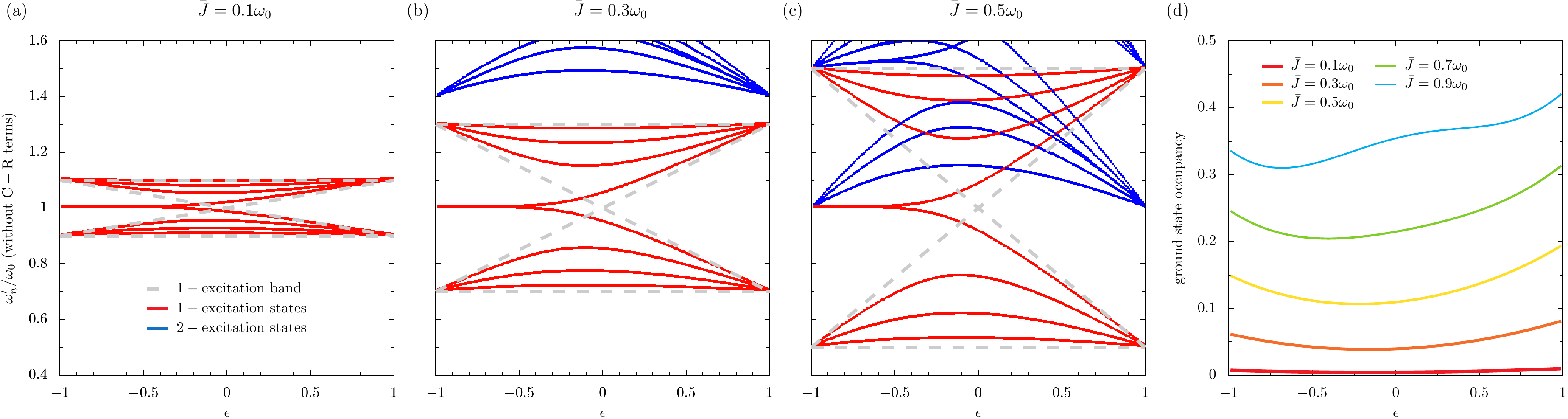}
 \caption{ \textbf{Effective band theory and renormalization of the vacuum.} Panels (a, b, c): the RWA eigenfrequencies for a finite chain $\omega_n^\prime$ as a function of the dimerization parameter $\epsilon$, with increasing coupling strength $\bar{J}$ in each column [cf. Eq.~\eqref{eq:xzzcxzczc}]. Dashed grey lines: the boundaries of the RWA eigenfrequencies in the one-excitation sector, as found from a periodic calculation in the continuum limit $N \to \infty$ [cf. Eq.~\eqref{eq:sddfsddadfsdf}]. Solid lines: one-excitation states are shown in red and two-excitation states are shown in blue. All data above $1.6\omega_0$ and below $0.4\omega_0$ is cut. Panel (d): The occupancy of the ground state of the dimerized chain as a function of $\epsilon$, where the $\bar{J}$ increases with increasing line thickness. In the figure, $N = 8$.}
 \label{INFgap}
\end{figure}

The preceding discussion of Fig.~\ref{pr33}~panels~(a-f) used terminology such as effective bandstructures, effective bands and effective band gaps. Such concepts may reasonably be invoked, since the finite system studied using the Hamiltonian of Eq.~\eqref{eq:hammyy} has a natural analogue in the continuum limit. When the number of 2LSs in the chain $N \to \infty$, the resultant eigenfrequencies may be grouped into certain bands which are separated by certain band gaps. For example, applying the RWA and concentrating on the first-excitation sector only, Eq.~\eqref{eq:hammyy} may be readily diagonalized as $\omega_0 \pm \sqrt{J_1^2 + J_2^2 + 2 J_1 J_2 \cos \left( q d\right) }$, where $q$ is the wavenumber and where periodic boundary conditions were employed to ensure an analytic result~\cite{ Asboth2016}. This simple expression exposes the presence of two bands ($\pm$ in the aforementioned expression) filled of one-excitation states, which appear inside the bow tie shape formed by
\begin{equation}
\label{eq:sddfsddadfsdf}
\omega_0 + \bar{J},
\quad\quad\quad\quad\quad\quad
 \omega_0 + |\epsilon| \bar{J},
 \quad\quad\quad\quad\quad\quad
 \omega_0 - |\epsilon| \bar{J},
 \quad\quad\quad\quad\quad\quad
 \omega_0 - \bar{J},
\end{equation} 
which are plotted as the dashed grey lines in Fig.~\ref{INFgap}~(a, b, c). We also plot the RWA results for a finite chain of size $N = 8$ in panels~(a, b, c), for increasing values of the coupling strength $\bar{J}$, which clearly fit the bow tie analysis of Eq.~\eqref{eq:sddfsddadfsdf} for the one-excitation eigenfrequencies (red lines). The two-excitation eigenfrequencies (blue lines) encroaching the bow tie for larger $\bar{J}$ also form a band, but no simple analytic expression for the two-excitation band is readily obtainable. Similar results to Fig.~\ref{INFgap}~(a, b, c) also hold with the C-R terms included, but since the number of excitations is no longer conserved the effective band theory language becomes less and less useful for larger values of $\bar{J}$. Notably, the edge states falling outside of the bow ties in Fig.~\ref{INFgap}~(a, b, c) are a priori excluded from the periodic boundary condition calculation leading to Eq.~\eqref{eq:sddfsddadfsdf}.

As was mentioned in the discussion of a 2LS dimer around Fig.~\ref{dimerr}~(c), ultrastrong coupling renormalizes the ground state due to the presence of the C-R terms in the Hamiltonian of Eq.~\eqref{eq:hammyy} linking the vacuum state and occupied states. This effect can be clearly seen for the dimerized chain by considering the occupancy of the ground state $\kettwo{\psi_1}$, as is shown in Fig.~\ref{INFgap}~(d) as a function of the dimerization $\epsilon$. The coupling strength $\bar{J}$ increases from $0.1\omega_0$ to $0.9\omega_0$ with decreasing line thickness. Quite intuitively, with smaller $\bar{J}$ (thick red line) the ground state is essentially unoccupied $\kettwo{\psi_1} = \ket{0, 0, ..., 0}$, being the trivial ground state familiar from the RWA. However, upon entering ultrastrong coupling (thinner lines), there is a significant occupation of the ground state $\kettwo{\psi_1}$. Importantly, this effect is highly sensitive to the dimerization $\epsilon$, and thus may be readily probed by taking measurements of the same physical system in different geometric configurations.\\


\noindent \textbf{Discussion}\\
We have studied a prototypical topological model in the ultrastrong coupling regime, where the rotating wave approximation breaks down. Inspired by the surge in experimental activity on one-dimensional dimerized arrays~\cite{Parto2018, Ota2018, Hadad2018, Han2019}, we have shown that in the ultrastrong coupling regime various desirable features remain. For example, topological edge states continue to form in the gaps in the effective bandstructure, and although they are no longer pinned at a constant `zero energy' they are highly localized in the topologically nontrivial geometric arrangement and indeed disappear in the trivial arrangement. Consideration of higher rungs of the energy ladder led to the discovery of anti-edge states, which have the property of residing everywhere apart from the edges of the chain. Finally, a hallmark of ultrastrong coupling is a non-empty ground state, and here we show that this vacuum renormalization is geometric-dependent for a dimerized chain, proffering opportunities for observation. We believe that our results should stimulate experimental work~\cite{Forn2010, Niemczyk2010, Forn2017, Mueller2020} in ultrastrong topology, as well as further theoretical work in this exciting area.\\


\noindent \textbf{Methods}\\
In this theoretical work, exact diagonalization of finite matrices was carried out as described in the main text. Further details are given in the Supplementary Information.\\

\noindent \textbf{Acknowledgments}\\
\textit{Funding}: CAD is supported by the Royal Society via a University Research Fellowship (URF\slash R1\slash 201158) and a Royal Society Research Grant (RGS\slash R1 \slash 211220). The latter grant also supported the summer internship of AJT. \textit{Discussions}: We thank T.~J.~Sturges for fruitful discussions.\\

\noindent \textbf{Author contributions}\\
CAD conceived of the study, performed the calculations and wrote the first version of the manuscript. AJT performed the numerical calculations and helped to draft the manuscript. Both authors gave final approval for publication and agree to be held accountable for the work performed therein.\\

\noindent \textbf{Competing interests}\\
The authors declare no competing interests.\\

\noindent \textbf{Data availability}\\
There is no additional data. Further information is given in the Supplementary Information.\\

\noindent \textbf{Additional information}\\
Supplementary Information is available for this paper.\\


\begin{thebibliography}{100}



\bibitem{Chang2018}
D.~E.~Chang, J.~S.~Douglas, A.~Gonzalez-Tudela, C.-L.~Hung and H.~J.~Kimble, 
Colloquium: Quantum matter built from nanoscopic lattices of atoms and photons,
\href{https://doi.org/10.1103/RevModPhys.90.031002}
{Rev. Mod. Phys. \textbf{90}, 031002 (2018)}.


\bibitem{Lu2014}
L.~Lu, J.~D.~Joannopoulos and M.~Soljacic,
Topological photonics,
\href{https://doi.org/10.1038/nphoton.2014.248}
{Nature Photon. \textbf{8}, 821 (2014)}.

\bibitem{Ozawa2019}
T.~Ozawa, H.~M.~Price, A.~Amo, N.~Goldman, M.~Hafezi, L.~Lu, M.~C.~Rechtsman, D.~Schuster, J.~Simon, O.~Zilberberg and I.~Carusotto,
Topological photonics,
\href{https://doi.org/10.1103/RevModPhys.91.015006}
{Rev. Mod. Phys. \textbf{91}, 015006 (2019)}.

\bibitem{Smirnova2020}
D.~Smirnova, D.~Leykam, Y.~Chong and Y.~Kivshar,
Nonlinear topological photonics,
\href{https://doi.org/10.1063/1.5142397}
{Appl. Phys. Rev. \textbf{7}, 021306 (2020)}.

\bibitem{McClarty2021}
P.~McClarty,
Topological magnons: a review,
\href{https://arxiv.org/abs/2106.01430}
{arXiv:2106.01430}.

\bibitem{Cooper2019}
N.~R.~Cooper, J.~Dalibard and I.~B.~Spielman,
Topological bands for ultracold atoms,
\href{https://doi.org/10.1103/RevModPhys.91.015005}
{Rev. Mod. Phys. \textbf{91}, 015005 (2019)}.


\bibitem{Mittal2018}
S.~Mittal, E.~A.~Goldschmidt and M.~Hafezi,
A topological source of quantum light,
\href{https://doi.org/10.1038/s41586-018-0478-3}
{Nature \textbf{561}, 502 (2018)}.


\bibitem{Blanco2018}
A.~Blanco-Redondo, B.~Bell, D.~Oren, B.~J.~Eggleton and M.~Segev,
Topological protection of biphoton states,
\href{https://doi.org/10.1126/science.aau4296}
{Science \textbf{362}, 568 (2018)}.


\bibitem{Bandres2018}
M.~A.~Bandres, S.~Wittek, G.~Harari, M.~Parto, J.~Ren, M.~Segev, D.~N.~Christodoulides and M.~Khajavikhan,
Topological insulator laser: Experiments,
\href{https://doi.org/10.1126/science.aar4005}
{Science \textbf{359}, eaar4005 (2018)}.

\bibitem{Barik2018}
S.~Barik, A.~Karasahin, C.~Flower, T.~Cai, H.~Miyake, W.~DeGottardi, M.~Hafezi and E.~Waks,
A topological quantum optics interface,
\href{https://doi.org/10.1126/science.aaq0327}
{Science \textbf{359}, 666 (2018)}.

\bibitem{Allen1975}
L.~Allen, and J.~H.~Eberly,
\textit{Optical Resonance and Two-Level Atoms} (Wiley, New York, 1975).

\bibitem{Kockum2019}
A.~F.~Kockum, A.~Miranowicz, S.~De~Liberato, S.~Savasta and F.~Nori,
Ultrastrong coupling between light and matter,
\href{https://doi.org/10.1038/s42254-018-0006-2}
{Nat. Rev. Phys. \textbf{1}, 19 (2019)}.


\bibitem{Forn2019}
P.~Forn-Diaz, L.~Lamata, E.~Rico, J.~Kono and E.~Solano,
Ultrastrong coupling regimes of light-matter interaction,
\href{https://doi.org/10.1103/RevModPhys.91.025005}
{Rev. Mod. Phys. \textbf{91}, 025005 (2019)}.

\bibitem{Boite2020}
A.~Le~Boite,
Theoretical methods for ultrastrong light-matter interactions,
\href{https://doi.org/10.1002/qute.201900140}
{Adv. Quantum Technol. \textbf{3}, 1900140 (2020)}.


\bibitem{Peropadre2013}
B.~Peropadre, D.~Zueco, D.~Porras and J.~J.~Garcia-Ripoll,
Nonequilibrium and nonperturbative dynamics of ultrastrong coupling in open lines,
\href{https://doi.org/10.1103/PhysRevLett.111.243602}
{Phys. Rev. Lett. \textbf{111}, 243602 (2013)}.

\bibitem{Naether2014}
U.~Naether, J.~J.~Garcia-Ripoll, J.~J.~Mazo and D.~Zueco,
Quantum chaos in an ultrastrongly coupled bosonic junction,
\href{https://doi.org/10.1103/PhysRevLett.112.074101}
{Phys. Rev. Lett. \textbf{112}, 074101 (2014)}.

\bibitem{Sanchez2014}
E.~Sanchez-Burillo, D.~Zueco, J.~J.~Garcia-Ripoll and L.~Martin-Moreno,
Scattering in the ultrastrong regime: nonlinear optics with one photon,
\href{https://doi.org/10.1103/PhysRevLett.113.263604}
{Phys. Rev. Lett. \textbf{113}, 263604 (2014)}.


\bibitem{Calvo2020}
J.~Calvo, D.~Zueco and L.~Martin-Moreno,
Ultrastrong coupling effects in molecular cavity QED,
\href{https://doi.org/10.1515/nanoph-2019-0403}
{Nanophotonics \textbf{9}, 277 (2020)}.

\bibitem{Roman2020}
J.~Roman-Roche, E.~Sanchez-Burillo and D.~Zueco,
Bound states in ultrastrong waveguide QED,
\href{https://doi.org/10.1103/PhysRevA.102.023702}
{Phys. Rev. A \textbf{102}, 023702 (2020)}.

\bibitem{Gonzalez2021}
C.~A.~Gonzalez-Gutierrez, J.~Roman-Roche and D.~Zueco,
Distant emitters in ultrastrong waveguide QED: ground-state properties and non-Markovian dynamics,
\href{https://doi.org/10.1103/PhysRevA.104.053701}
{Phys. Rev. A \textbf{104}, 053701 (2021)}.

\bibitem{Asboth2016}
J.~K.~Asboth, L.~Oroszlany and A.~Palyi,
\textit{A Short Course on Topological Insulators} (Springer, Heidelberg, 2016).

\bibitem{Haldane2017}
F.~D.~M.~Haldane,
Nobel lecture: topological quantum matter,
\href{https://doi.org/10.1103/RevModPhys.89.040502}
{Rev. Mod. Phys. \textbf{89}, 040502 (2017)}.

\bibitem{Forn2010}
P.~Forn-Diaz, J.~Lisenfeld, D.~Marcos, J.~J.~Garcia-Ripoll, E.~Solano, C.~J.~P.~M.~Harmans and J.~E.~Mooij,
Observation of the Bloch-Siegert shift in a qubit-oscillator system in the ultrastrong coupling regime,
\href{https://doi.org/10.1103/PhysRevLett.105.237001}
{Phys. Rev. Lett. \textbf{105}, 237001 (2010)}.

\bibitem{Niemczyk2010}
T.~Niemczyk, F.~Deppe, H.~Huebl, E.~P.~Menzel, F.~Hocke, M.~J.~Schwarz, J.~J.~Garcia-Ripoll, D.~Zueco, T.~Hummer, E.~Solano, A.~Marx and R.~Gross, 
Circuit quantum electrodynamics in the ultrastrong-coupling regime,
\href{https://doi.org/10.1038/nphys1730}
{Nat. Phys. \textbf{6}, 772 (2010)}.

\bibitem{Forn2017}
P.~Forn-Diaz, J.~J.~Garcia-Ripoll, B.~Peropadre, J.-L.~Orgiazzi, M.~A.~Yurtalan, R.~Belyansky, C.~M.~Wilson and A.~Lupascu, 
Ultrastrong coupling of a single artificial atom to an electromagnetic continuum in the nonperturbative regime,
\href{https://doi.org/10.1038/nphys3905}
{Nat. Phys. \textbf{13}, 39 (2017)}.



\bibitem{Mueller2020}
N.~S.~Mueller, Y.~Okamura, B.~G.~M.~Vieira, S.~Juergensen, H.~Lange, E.~B.~Barros, F.~Schulz and S.~Reich, 
Deep strong light-matter coupling in plasmonic nanoparticle crystals,
\href{https://doi.org/10.1038/s41586-020-2508-1}
{Nature \textbf{583}, 780 (2020)}.

\bibitem{Liu2021}
S.~Liu, Z.~Lu and H.~Zheng, 
Geometric phase and non-adiabatic resonance of the Rabi model,
\href{https://doi.org/10.1088/1751-8121/ac2a04}
{J. Phys. A: Math. Theor. \textbf{54}, 445302 (2021)}.



\bibitem{Ying2022}
Z.-J.~Ying, 
From quantum Rabi model to Jaynes–Cummings model: Symmetry-breaking quantum phase transitions, symmetry-protected topological transitions and multicriticality,
\href{https://doi.org/10.1002/qute.202100088}
{Adv. Quantum Technol. \textbf{5}, 2100088 (2022)}.

\bibitem{Ficek2002}
Z.~Ficek and R.~Tanas, 
Entangled states and collective nonclassical effects in two-atom systems,
\href{https://doi.org/10.1016/S0370-1573(02)00368-X}
{Phys. Rep. \textbf{372}, 369 (2002)}.

\bibitem{Downing2019}
C.~A.~Downing, J.~C.~L\'{o}pez~Carre\~{n}o, F.~P.~Laussy, E.~del Valle and A.~I.~Fern\'{a}ndez-Dom\'{i}nguez,
Quasichiral interactions between quantum emitters at the nanoscale,
\href{https://doi.org/10.1103/PhysRevLett.122.057401}
{Phys. Rev. Lett. \textbf{122}, 057401 (2019)}.


\bibitem{Downing2020}
C.~A.~Downing, J.~C.~L\'{o}pez~Carre\~{n}o, E.~del Valle and A.~I.~Fern\'{a}ndez-Dom\'{i}nguez,
Asymmetric coupling between two quantum emitters,
\href{https://doi.org/10.1103/PhysRevA.102.013723}
{Phys. Rev. A \textbf{102}, 013723 (2020)}.


\bibitem{Decordi2017}
G.~L.~Decordi and A.~Vidiella-Barranco,
Two coupled qubits interacting with a thermal bath: A comparative study of different models,
\href{https://doi.org/10.1016/j.optcom.2016.10.017}
{Opt. Commun. \textbf{387}, 366 (2017)}.





\bibitem{Su1979}
W.~P.~Su, J.~R.~Schrieffer and A.~J.~Heeger,
Solitons in polyacetylene,
\href{https://doi.org/10.1103/PhysRevLett.42.1698}
{Phys. Rev. Lett. \textbf{42}, 1698 (1979)}.


\bibitem{Bell1970}
R.~J.~Bell and P.~Dean,
Atomic vibrations in vitreous silica,
\href{https://doi.org/10.1039/DF9705000055}
{Discuss. Faraday Soc. \textbf{50}, 55 (1970)}.

\bibitem{Thouless1974}
D.~J.~Thouless,
Electrons in disordered systems and the theory of localization,
\href{https://doi.org/10.1016/0370-1573(74)90029-5}
{Phys. Rep. \textbf{13}, 93 (1974)}.

\bibitem{DowningMartin}
C.~A.~Downing and L.~Mart\'{i}n-Moreno,
Polaritonic Tamm states induced by cavity photons,
\href{https://doi.org/10.1515/nanoph-2020-0370}
{Nanophotonics \textbf{10}, 513 (2021)}.

\bibitem{Pablo}
P.~Martinez Azcona and C.~A.~Downing,
Doublons, topology and interactions in a one-dimensional lattice,
\href{https://doi.org/10.1038/s41598-021-91778-z}
{Sci Rep \textbf{11}, 12540  (2021)}.

\bibitem{Parto2018}
M.~Parto, S.~Wittek, H.~Hodaei, G.~Harari, M.~A.~Bandres, J.~Ren, M.~C.~Rechtsman, M.~Segev, D.~N.~Christodoulides and M.~Khajavikhan,
Edge-mode lasing in 1D topological active arrays,
\href{https://doi.org/10.1103/PhysRevLett.120.113901}
{Phys. Rev. Lett. \textbf{120}, 113901 (2018)}.

\bibitem{Ota2018}
Y.~Ota, R.~Katsumi, K.~Watanabe, S.~Iwamoto and Y.~Arakawa,
Topological photonic crystal nanocavity laser,
\href{https://doi.org/10.1038/s42005-018-0083-7}
{Commun. Phys. \textbf{1}, 86 (2018)}.

\bibitem{Hadad2018}
Y.~Hadad, J.~C.~Soric, A.~B.~Khanikaev and A.~Alu,
Self-induced topological protection in nonlinear circuit arrays,
\href{https://doi.org/10.1038/s41928-018-0042-z}
{Nat. Electron. \textbf{1}, 178 (2018)}.

\bibitem{Han2019}
C.~Han, M.~Lee, S.~Callard, C.~Seassal and H.~Jeon,
Lasing at topological edge states in a photonic crystal L3 nanocavity dimer array,
\href{https://doi.org/10.1038/s41377-019-0149-7}
{Light Sci. Appl. \textbf{8}, 40 (2019)}.


\end{thebibliography}
\end{document}